# A General Class of Metamaterial Transformation Slabs

Ilaria Gallina [1], Giuseppe Castaldi [1], Vincenzo Galdi [1], Andrea Alù [2], and Nader Engheta [3]

(1) Waves Group, Department of Engineering, University of Sannio, I-82100, Benevento, Italy

(2) Department of Electrical and Computer Engineering, The University of Texas at Austin, Austin, TX 78712, USA

(3) Department of Electrical and Systems Engineering, University of Pennsylvania, Philadelphia, PA 19104, USA

## Abstract

In this paper, we apply transformation-based optics to the derivation of a general class of transparent metamaterial slabs. By means of analytical and numerical full-wave studies, we explore their image displacement/formation capabilities, and establish intriguing connections with configurations already known in the literature. Starting from these revisitations, we develop a number of nontrivial extensions, and illustrate their possible applications to the design of perfect radomes, anti-cloaking devices, and focusing devices based on double-positive (possibly nonmagnetic) media. These designs show that such anomalous features may be achieved without necessarily relying on negative-index or strongly resonant metamaterials, suggesting more practical venues for the realization of these devices.

## 1. Introduction

First envisaged in the 1960s within the framework of electromagnetic (EM) wave propagation in curved space-time or curved-coordinate systems (see, e.g., Refs. 1 and 2), "transformation optics" (TO) [3-7] (also referred to as "transformation EM") has only recently become a technologically viable approach to EM field manipulation, thanks to the formidable advances in the synthesis of "metamaterials" with precisely controllable anisotropy and spatial-inhomogeneity properties (see, e.g., Ref. 8).

In essence, TO relies on the formal invariance of Maxwell equations under coordinate transformations, which enables the design of the desired field behavior in a *curved-coordinate* fictitious space, and its subsequent translation in a *flat-metric* physical space filled up by a suitable (anisotropic, spatially-inhomogeneous) "transformation medium" which embeds the coordinate-transformation effects. As a sparse sample of the available application examples, besides the celebrated invisibility "cloaking" (see, e.g., Refs. 8 and 9), we recall here those pertaining to super/hyper-lensing, [10-12] field concentrators [13] and rotators, [14] conformal sources, [15] beam shifters and splitters, [16,17] retroreflectors, [18,19] as well as the broad framework of "illusion optics" [20] (see also Ref. 21, and references therein, for a collection of recent applications).

In this paper, we focus on a general class of TO-inspired metamaterial slabs (henceforth, simply referred to as "transformation slabs"). Propagation of EM waves in slab-type configurations is a subject of longstanding interest, which has recently received renewed attention in view of the new perspectives and degrees of freedom endowed by metamaterials. In particular, low-refractive-index slabs have been explored in connection with ultrarefraction phenomena, with potential applications to directive emission, [22] enhanced transmission through subwavelength apertures, [23] and phase-pattern tailoring. [24] *Double-negative* (DNG) slabs have attracted a great deal of attention, especially in connection with superlensing [25] and phase-compensation [26] effects. Superlensing effects have also been predicted in metamaterial slabs featuring *extreme* effective parameters, [27–29] whereas anomalous tunneling and transparency phenomena have been studied in connection with *single-negative* (SNG) bi-layers made of *epsilon-negative* (ENG) and *mu-negative* (MNG) media. [30] Some of the above effects, attributable to anomalous interactions occurring in DNG/double-positive (DPS) or ENG/MNG paired configurations, have also been interpreted within the broader context of "complementary" media. [31]

Within this framework, TO has been shown to provide insightful alternative interpretations, and to enable for nontrivial extensions (see, e.g., Refs. 10 and 32–34 for a sparse sampling). For instance, lensing configurations have been interpreted in terms of *coordinate-folding*, [10] and have been extended so as to deal with the presence of passive or active objects. [34] More in general, there has been considerable recent interest in the study of transparency/reflection conditions [35–37] in scenarios involving transformation media.

Against this background, in this paper, we introduce and explore a general class of two-dimensional (2-D) coordinate transformations capable of yielding *transparent* metamaterial slabs potentially useful for image displacement and reconstruction.

First, we show that the proposed class of transformation slabs includes as particular cases several conventional configurations already known in the literature. Next, we focus on a series of nontrivial extensions, with possible applications to radome, cloaking/anti-cloaking, and focusing scenarios, which naturally emerge from the proposed approach. Our results provide further confirmation of the broad breadth of TO and its intriguing potentials as a general *unifying approach* to metamaterial functionalization. Our interest is especially focused in proposing designs that may not require negative constitutive parameters, but that exploit the inherent anisotropy of the transformation slabs to achieve anomalous wave interaction within the DPS regime of operation. This may be particularly attractive to simplify the realization of these devices and to possibly relax some bandwidth limitations typical of SNG and DNG metamaterials.

Accordingly, the rest of the paper is laid out as follows. In Sec. 2, we outline the problem geometry and its formulation. In Sec. 3, we illustrate the general analytical solution and some details about the computational tools utilized. In Sec. 4, we focus on particular reductions corresponding to configurations already known in the literature. In Sec. 5, starting from these configurations, we present a number of nontrivial extensions, with applications to perfect radomes, cloaking/anti-cloaking, and focusing scenarios. Finally, in Sec. 6, we provide some brief conclusions together with hints for future research.

## 2. Problem Geometry and Formulation

Referring to the geometry in Fig. 1, we consider a 2-D coordinate transformation between a fictitious (vacuum) space $(x', y', z')$ and the slab region $|x| \leq d$ (embedded in vacuum) in the actual physical space $(x, y, z)$,

$$\begin{cases} x' = au(x), \\ y' = \dfrac{y}{\dot{u}(x)} + v(x), \\ z' = z, \end{cases} \quad (1)$$

where $a$ is a real scaling parameter, and $u(x)$ and $v(x)$ are arbitrary continuous real functions. In (1) and henceforth, an overdot denotes differentiation with respect to the argument. We assume that the derivative $\dot{u}(x)$ is continuous and nonvanishing within the slab region, so that the coordinate transformation in (1) is likewise continuous. Moreover, for reasons that will become clearer hereafter, we also assume that

$$\dot{u}(\pm d) = 1, \quad (2)$$

so that, apart from a possible (irrelevant) rigid translation, the planar interfaces $x' = d'_{1,2} \equiv au(\mp d)$ of the (virtual) vacuum slab in the auxiliary space are imaged as the planar interfaces $x = \mp d$ of the transformed slab region in the actual physical space (see Fig. 1). Within the TO framework [3-7], the EM field effects induced by the curved metrics in (1) can be equivalently obtained in a flat, Cartesian space by filling-up the transformed slab region $|x| \le d$ with an anisotropic, spatially inhomogenous transformation medium described by the relative permittivity and permeability tensors:

$$\underline{\underline{\varepsilon}}(x, y) = \underline{\underline{\mu}}(x, y) = \det\left[\underline{\underline{J}}(x, y)\right] \underline{\underline{J}}^{-1}(x, y) \cdot \left[\underline{\underline{J}}^{-1}(x, y)\right]^T, \quad (3)$$

where the superscript $T$ denotes matrix transposition, and

$$\underline{\underline{J}}(x, y) = \frac{\partial(x', y', z')}{\partial(x, y, z)} = \begin{bmatrix} a\dot{u}(x) & 0 & 0 \\ \dot{v}(x) - \dfrac{\ddot{u}(x) y}{\dot{u}^2(x)} & \dfrac{1}{\dot{u}(x)} & 0 \\ 0 & 0 & 1 \end{bmatrix} \quad (4)$$

is the Jacobian matrix. It can be shown (see Appendix A for details) that the constitutive tensors in (4) are *real* and *symmetric*, and hence they admit real eigenvalues, and they can be diagonalized by an orthogonal matrix. In particular, it is expedient to represent these tensors in their diagonalized forms ($\underline{\underline{\tilde{\varepsilon}}}$ and $\underline{\underline{\tilde{\mu}}}$, respectively) in the principal reference system $(\xi, \upsilon, z)$ constituted by their (orthogonal) eigenvectors, viz.,

$$\underline{\underline{\tilde{\varepsilon}}}(x,y) = \underline{\underline{\tilde{\mu}}}(x,y) = \begin{bmatrix} \Lambda_\xi(x,y) & 0 & 0 \\ 0 & \Lambda_\upsilon(x,y) & 0 \\ 0 & 0 & a \end{bmatrix}, \qquad (5)$$

where $\Lambda_{\xi,\upsilon}$ denote the transverse eigenvalues. It can be shown (see Appendix A for details) that

$$\text{sgn}\left[\Lambda_{\xi,\upsilon}(x,y)\right] = \text{sgn}(a), \qquad (6)$$

and hence, depending on the sign of the scaling parameter $a$, the constitutive tensors are either both *positive-defined* $(a>0)$ or both *negative-defined* $(a<0)$. Accordingly, the resulting transformation medium is either DPS $(a>0)$ or DNG $(a<0)$. Moreover, it readily follows from (5) that, for the particular choice $a=1$, the resulting transformation medium is effectively *non-magnetic* $(\tilde{\mu}_z = 1)$ for transverse-magnetic (TM) polarization (i.e., $z$-directed magnetic field), and *non-electric* $(\tilde{\varepsilon}_z = 1)$ for transverse-electric (TE) polarization (i.e., $z$-directed electric field).

Finally, recalling the developments in Ref. 36, we note that the condition (2) (i.e., the fact that at the transformed-region boundaries $x = \mp d$ the coordinate mapping in (1) reduces to a rigid translation) ensures the *reflectionless* of the transformation slab.

In what follows, without loss of generality, we study the EM response of the above defined transformation slab excited by a time-harmonic $(\exp(-i\omega t))$, TM-polarized, arbitrary aperture field distribution located at a source-plane $x = x_s < -d$,

$$H_z(x = x_s, y) = f(y). \qquad (7)$$

## 3. Generalities on Analytical Derivations and Numerical Simulations

*3.1 General solution*

It can be shown (see Appendix B for details) that, for an observer located at $x_o > d$ (i.e., beyond the transformation slab), the EM response of the system can be viewed as effectively generated by an equivalent aperture field distribution

$$H_z(x = x_i, y) = f(y - y_0), \quad (8)$$

corresponding to a (virtual or real) image at the plane

$$x = x_i \equiv x_s + 2d - a[u(d) - u(-d)], \quad (9)$$

which exactly reproduces the source-plane aperture field distribution in (7), apart from a possible rigid translation of

$$y_0 = v(-d) - v(d) \quad (10)$$

along the *y*-axis. Note that the image-plane position and the *y*-displacement *do not* depend on the actual form of the mapping function $u(x)$ and $v(x)$ in (1), but rather on their values at the slab interfaces $x = \mp d$, thereby leaving ample design freedom. Clearly, for $x_i < d$, a *virtual* image is formed, corresponding to a perceived displacement of the source. In particular, for $x_i = x_s$ and $y_0 = 0$ (i.e., zero displacement), the transformation slab behaves like a vacuum slab of same thickness, thereby acting as a *perfect radome*. Conversely, for $x_i > d$, a *real* image is formed, and the slab ideally behaves like a perfect lens.

*3.2 Simulation tools and parameters*

Our analytical derivations below are supported by full-wave numerical studies based on a finite-element solution of the EM problem for Gaussian-beam or plane-wave illuminations. Our simulations rely on the use of the commercial software COMSOL Multiphysics, [38] based on the finite-element method (FEM), in view of its ability to deal with arbitrary anisotropic and spatially inhomogeneous transformation media. In all simulations below, slight material losses (loss-tangent=$10^{-3}$) are assumed, and the slabs are truncated along the *y*-axis to an aperture of $2L$. The resulting computational domains are adaptively discretized via nonuniform meshing (with element size which can be locally as small as $\sim 4\times 10^{-4} \lambda_0$, with $\lambda_0$ denoting the vacuum wavelength), and terminated by perfectly matched layers, resulting into a total number of unknowns $\sim 10^6$.

## 4. Particular Cases Already Known in the Literature

In order to illustrate the general character of the proposed class, we begin showing that it includes as special cases several configurations already known in the literature.

Perhaps the most trivial reduction is obtained by assuming $u(x) = x, v(x) = 0$, which yields

$$\underline{\underline{\varepsilon}} = \underline{\underline{\mu}} = \begin{bmatrix} a^{-1} & 0 & 0 \\ 0 & a & 0 \\ 0 & 0 & a \end{bmatrix}, \quad (11)$$

which closely resembles the class of nonreflecting birefringent metamaterials considered in Ref. 39 (in connection with perfect lensing for vectorial fields), and also widely used in the design of perfectly-matched layers for finite-difference-time-domain numerical schemes (see, e.g., Ref. 40). In this case, (9) and (10) trivially reduce to

$$x_i = x_s + 2d(1-a), \quad y_0 = 0, \quad (12)$$

corresponding to an *orthogonal* (with respect to the slab) image displacement, which vanishes only for $a = 1$ (i.e., when the material trivially reduces to vacuum), and yields Pendry's perfect lens [22] for $a = -1$.

Another interesting reduction is obtained by assuming

$$a = 1, \quad u(x) = x, \quad v(x) = \alpha x, \quad (13)$$

which yields

$$\underline{\underline{\varepsilon}} = \underline{\underline{\mu}} = \begin{bmatrix} 1 & -\alpha & 0 \\ -\alpha & 1+\alpha^2 & 0 \\ 0 & 0 & 1 \end{bmatrix}. \quad (14)$$

It is interesting to note that the above medium (spatially homogeneous, DPS, and non-magnetic for the assumed TM polarization) is amenable to the *tilted uniaxial* class considered in Ref. 41. Indeed, letting

$$\alpha = \left( \varepsilon_R - \frac{1}{\varepsilon_R} \right) \sin\theta_R \cos\theta_R, \tag{15}$$

and enforcing the condition for *omnidirectional total transmission* in Eq. (30) of Ref. 41 (assuming $\varepsilon_L = 1$),

$$\varepsilon_R \sin^2\theta_R + \frac{1}{\varepsilon_R} \cos^2\theta_R = 1, \tag{16}$$

the diagonalized constitutive tensors [cf. (5)] in the principal reference system (rotated of an angle $\theta_R$ with respect of the *x*-axis) become

$$\tilde{\underline{\underline{\varepsilon}}} = \tilde{\underline{\underline{\mu}}} = \begin{bmatrix} \varepsilon_R & 0 & 0 \\ 0 & \varepsilon_R^{-1} & 0 \\ 0 & 0 & 1 \end{bmatrix}, \tag{17}$$

and hence completely equivalent (at least for the assumed TM polarization) to that considered in Eq. (1) of Ref. 41 (taking into account the different reference system utilized). Under these assumptions, (9) and (10) reduce to

$$x_i = x_s, \quad y_0 = -2\alpha d, \tag{18}$$

corresponding to a lateral shift which can be tailored in sign and amplitude by tweaking the slab thickness and constitutive parameters [cf. (15)]. Note that, depending on the incident field and on the constitutive parameters, the above shift can be induced by *positive* or *negative* refraction (see Ref. 41 for details). As an example, Fig. 2 shows a typical FEM-computed response for a normally-incident collimated Gaussian beam, which illustrates the lateral shift induced for $\alpha > 0$. It is worth pointing out that the coordinate transformation in (13) was also proposed in Ref. 17 for the very purpose of beam shifting, without, however, establishing the connection with the tilted uniaxial media in (17) and Ref. 41.

As a final example of well-known configurations included as special cases in our proposed class, we consider the *real-image* case $x_i > d$ in (9). In this case, assuming $u(d) > u(-d)$ and recalling that $x_s < -d$, it readily follows from (9) that $a < 0$, and thus the slab is DNG. Accordingly, this case belongs to the broad class of DPS-DNG *complementary media* considered in Ref. 31, and its further generalizations (see, e.g., Ref. 34).

## 5. Examples of Nontrivial Extensions

*5.1 DPS nonmagnetic radomes*

As a first example of nontrivial extensions of the above illustrated configurations, we consider a general class of transparent slabs characterized by

$$a = 1, \quad u(x) = x, \quad v(d) = v(-d), \tag{19}$$

for which the image displacement is zero, viz.,

$$x_i = x_s, \quad y_0 = 0. \tag{20}$$

Such slabs exhibit the same EM response of a slab of vacuum of same thickness, and can therefore be viewed as DPS nonmagnetic "perfect radomes." The possibly simplest conceivable realization can be obtained by choosing

$$v(x) = \alpha |x|, \tag{21}$$

which, by comparison with (13), is readily realized to correspond to the juxtaposition of two beam shifters (as in Fig. 2) producing *opposite* lateral shifts. Figure 3 clearly illustrates the shift-compensation effect that yields the same transmitted field that a vacuum slab of same size would produce. Note that, at the interface $x = 0$ separating the two half-slabs, a *negative* (total) refraction takes place. Interestingly, a similar "twin-crystal" configuration (involving halfspaces instead of slabs) was also studied in Ref. 42 in connection with total *amphoteric* refraction. Again, a similar configuration was proposed in Ref. 17, in a beam-shifting framework, without establishing the connection with the twin-crystal case. Our extension above indicates that twin-crystal slabs are potentially interesting candidates for radome applications, in view of their relatively simple constitutive properties (which involve only piecewise anisotropic, homogeneous media, with everywhere finite parameters).

Note that (19) constrains only the boundary values of the function $v(x)$, and thus different choices are possible in principle. For instance, choosing

$$v(x) = \beta x^2 \tag{22}$$

yields an inhomogenous transformation medium, whose constitutive parameters are shown in Fig. 4. It can be observed that this medium still resembles the twin-crystal structure above, but with a spatial tapering in the constitutive parameters and in the optical-axis direction. The corresponding response is shown in Fig. 5, from which a phenomenon resembling the beam-shift compensation in Fig. 3 is still visible.

An alternative perfect-radome condition that does not rely on (global or local) crystal-twinning is given by

$$a = 1, \quad u(d) - u(-d) = 2d, \quad v(x) = 0, \tag{23}$$

with emphasis placed on the function $u(x)$, instead of $v(x)$. For instance, the class of functions satisfying the conditions

$$\dot{u}(x) = u_0 + u_1 \left(\frac{x}{d}\right)^n + u_2 \left(\frac{x}{d}\right)^{2n}, \tag{24}$$

with $n$ even, and

$$\begin{cases} u_0 + u_1 + u_2 = 1, \\ u_1 = -2u_2 \dfrac{1+n}{1+2n}, \end{cases} \tag{25}$$

provides a possible solution. It can be shown that, unlike the previous cases, the corresponding constitutive parameters tend to assume *extreme* values for $|y| \to \infty$. However, in view of the unavoidable slab truncation, this does not constitute a serious practical limitation. Figure 6 illustrate typical constitutive parameters for this configuration, from which the different structure (as compared to the twin-crystal-like cases) as well as the tendency towards extreme values are clearly visible. From the corresponding EM response, shown in Fig. 7, it can be observed that the beam profile is not subject to any lateral shift inside the slab.

Another interesting class is obtained by choosing $a = 1$, $v(x) = 0$, and

$$u(x) = \pm \left(\frac{1}{d+\delta}\right) \sqrt{\left[(\pm x + \delta)^2 - \delta^2 + \Delta^2\right]\left[(d+\delta)^2 - \delta^2 + \Delta^2\right]}, \quad x \gtrless 0, \tag{26}$$

where $\delta \geq 0$ is an offset parameter, while $\Delta$ is a small parameter that is eventually let tend to zero. Note that for $\delta = 0$ (and $\Delta \to 0$) the transformation trivially reduces to the identity, while for $\delta > 0$ (and $\Delta \to 0$) the mapping in (26) vanishes at the $x = 0$ plane; in this latter case, it can be shown that the constitutive parameters tend to exhibit extreme values at the $x = 0$ plane as well as for $|y| \to \infty$. Unlike the transformation slabs obeying to (19) or (23), which ideally behave as vacuum slabs of same thickness, the above configuration, while still being ideally nonreflecting, induces a nonzero image displacement along the $x$-axis [cf. (9)]

$$x_i - x_s = 2d - \left(\frac{2}{d+\delta}\right)\left[(d+\delta)^2 - \delta^2\right]. \tag{27}$$

Figure 8 shows representative constitutive parameter maps, from which the anticipated singular behavior at the $x = 0$ plane, as well as for $|y| \to \infty$, can be observed. The corresponding Gaussian-beam response, shown in Fig. 9, illustrates (by comparison with Figs. 3, 5, and 7) the image displacement, with the slight imperfect transmission and reflection effects attributable to the structure and material parameter truncations.

*5.2 DPS cloak/anti-cloak*

The last example of radome class is particularly interesting because the underlying transformations in (26) (intended for each of the half slabs) are directly related to that used in Ref. 13 for designing an invisibility cloak with *square* shape. The above results therefore seem to provide the building block for an "anti-cloaking" device, like those proposed in Refs. 43 and 44, in connection with cylindrical geometries. Such a device would be capable of allowing field penetration inside the cloak shell. However, unlike the designs proposed in Refs. 43 and 44 (based on DNG or SNG media, possibly exhibiting extreme-value parameters), it would involve *only DPS media*, thereby removing the most significant technological limitations that prevent their practical realization and that would cause significant bandwidth restrictions and loss mechanism.

As an illustrative example, we consider the configuration in Fig. 10(a), where four modified (trapezoidal-shaped, translated and possibly rotated) versions of the transformation slab in (26) are juxtaposed so as to form a square shell. Referring to the rightmost trapezoidal slab (the other three being simply obtained via translation and/or rotation), the underlying coordinate transformation entails $v(x) = 0$ and

$$u(x) = \begin{cases} -\dfrac{1}{x_1}\sqrt{(x^2 - x_2^2 - \Delta_2^2)(x_1^2 - x_2^2 - \Delta_2^2)}, & x_1 \leq x \leq x_2, |y| \leq x, \\ \dfrac{1}{x_3}\sqrt{(x^2 - x_2^2 + \Delta_3^2)(x_3^2 - x_2^2 + \Delta_3^2)}, & x_2 \leq x \leq x_3, |y| \leq x, \end{cases} \quad (28)$$

$$a = \begin{cases} -\dfrac{x_1^2}{x_1^2 - x_2^2 - \Delta_2^2}, & x_1 \leq x \leq x_2, |y| \leq x, \\ \dfrac{x_3^2}{x_3^2 - x_2^2 + \Delta_3^2}, & x_2 \leq x \leq x_3, |y| \leq x, \end{cases} \quad (29)$$

with $\Delta_{2,3}$ denoting vanishingly small parameters. At variance with the original transformation slab in (26), note the two different values of $a$ in (29), both positive (thereby indicating that the media are still DPS), but $\neq 1$ (thereby indicating that the media exhibit magnetic properties, as required in perfect cloaking/anti-cloaking transformations). In spite of the different coordinate-transformation and notation [45] that we utilize, the outer square shell in Fig. 10(a) shares the same spirit as the square cloak in Ref. 13; in both configurations, in the ideal limit ($\Delta_3 \to 0$ in our case), a point (coordinate-system origin) in the fictitious space is mapped into a square [of sidelength $2x_2$ in Fig. 10(a)] in the physical space, thereby creating a square "hole" effectively inaccessible by the EM fields (see, e.g., Fig. 1 in Ref. 13).

Figure 10(b) shows the response for oblique plane-wave incidence in the presence of the outer (cloak) shell only, from which one observes how the impinging radiation is re-routed, with little exterior scattering and interior penetration (attributed to the inevitable parameter truncations). Conversely, as shown in Fig. 10(c), the addition of the inner (anti-cloak) shell renders field penetration possible, with the restoration of a modal field inside the inner square region, in a fashion that closely resembles the interactions attainable via a SNG or DNG anti-cloak. [44] The corresponding material parameters are not shown explicitly, but their qualitative behaviors may be gathered from the infinite-slab configuration in Fig. 8. In particular, in the ideal limit $\Delta_{2,3} \to 0$, a singular behavior is expected at the interface separating the cloak and anti-cloak shells.

The above illustrated mechanisms also offer a suggestive idea for *partial* (i.e., angle-selective) cloaking. For instance, Fig. 11 shows the response of the cloak/anti-cloak square configuration above when two trapezoidal anti-cloak elements are removed. It can be observed that, when the illumination direction is orthogonal to the sides that are missing the anti-cloak elements [Fig. 11(a)], the cloaking mechanism prevails, and there is still little field penetration. Conversely, strong field penetration is obtained when the illumination direction is orthogonal to the sides corresponding to the two remaining anti-cloak elements [Fig. 11(b)].

*5.3 DPS focusing devices*

It was shown in Sec. 4 that the class of transformation slabs under analysis includes as special cases the perfect lenses based on DNG media. However, from (9), it may appear that a *perfect real image* (i.e., $x_i > d$) can also be formed by a *DPS transformation slab* (i.e., $a > 0$), provided that $u(d) < u(-d)$. Note that such condition, coupled with the reflectionless condition (2), implies that the derivative $\dot{u}(x)$ must vanish somewhere inside the slab, thereby yielding extreme-value constitutive parameters and extreme anisotropy. It should be pointed out that under this extreme condition, the previous analytical derivations (based on the continuity of the coordinate transformation) break down, and so the seemingly implied "perfect" imaging and phase compensation are not strictly attainable. Nevertheless, one can still obtain some interesting focusing effects, which bear a resemblance with those observed in Refs. 27─29 in connection with anisotropic media exhibiting extreme-value parameters along specific directions.

As an example, assuming $x_i = -x_s > d$, it can easily be verified that a transformation featuring $a = 1$, $v(x) = 0$, and

$$u(x) = x\left(1 - 3\frac{x_i \pm \Delta}{2d}\right) + \frac{x_i \pm \Delta}{2d^3}x^3, \quad x \gtrless x_p \equiv \sqrt{\frac{2d^3}{3x_i}\left(\frac{3x_i}{2d} - 1\right)}, \qquad (30)$$

with $\Delta$ denoting a vanishingly small parameter, would satisfy the above conditions. Figure 12 shows some representative constitutive parameter maps, from which the singular behavior at the planes $x = \pm x_p$ [where the derivative of $u(x)$ in (30) vanishes] is evident. The corresponding EM response for a subwavelength-waisted Gaussian beam excitation, shown in Fig. 13, illustrates the achievable focusing effects.

*5.4 Possible further twists: SNG media via complex mapping*

From Eq. (6), it is clear that the transformation media in (3) can only be DPS or DNG, but not SNG. Interestingly, SNG media can be obtained via a *complex-coordinate* modification of the transformation in (1), viz.,

$$\begin{cases} x' = iau(x), \\ y' = i\dfrac{y}{\dot{u}(x)} + iv(x), \\ z' = z. \end{cases} \qquad (31)$$

In this case, via straightforward extension of the results in Appendix A, it can be shown that the constitutive tensors in the principal reference system assume the form

$$\underline{\underline{\tilde{\varepsilon}}}(x,y) = \underline{\underline{\tilde{\mu}}}(x,y) = \begin{bmatrix} \Lambda_\xi(x,y) & 0 & 0 \\ 0 & \Lambda_\upsilon(x,y) & 0 \\ 0 & 0 & -a \end{bmatrix}, \qquad (32)$$

while (6) remains still valid. Accordingly, the resulting medium is SNG. Note that the framework utilized in Sec. 3.1 cannot be directly applied to the study of configurations involving paired ENG/MNG transformation media, since the coordinate mapping needs to be continuous. However, more or less straightforward extensions may cover these cases too.

## 6. Conclusions and Perspectives

In this paper, we have studied a general class of TO-based transparent metamaterial slabs. Via analytical derivations and numerical-full wave simulations, we have explored the image displacement and reconstruction capabilities of such slabs, highlighting the wide breadth of phenomenologies involved. Moreover, starting from those special cases corresponding to configurations already known in the literature, we have developed some nontrivial extensions, and illustrated their potential applications to the design of radomes, cloaking and anti-cloaking devices, and focusing devices based only on DPS anisotropic media.

Our results confirm the power of TO as a systematic approach to the design of application-oriented metamaterials with prescribed (e.g., DPS, nonmagnetic) constitutive properties.

It should be noted that no attempt was made at this stage to optimize the parametric configurations, and emphasis was only placed on the illustration of the basic phenomenologies. Parametric optimization, as well as deeper exploration of certain interactions (especially in connection with the DPS anti-cloaking and focusing), are currently being pursued. Also worth of interest is the extension to ENG/MNG paired configurations.

**Appendix A: Pertaining to (5) and (6)**

Substituting (4) into (3) yields

$$\underline{\underline{\varepsilon}}_r(x,y) = \underline{\underline{\mu}}_r(x,y) = \begin{bmatrix} \dfrac{1}{a\dot{u}^2(x)} & \dfrac{-\dot{u}^2(x)\dot{v}(x) + y\ddot{u}(x)}{a\dot{u}^3(x)} & 0 \\ \dfrac{-\dot{u}^2(x)\dot{v}(x) + y\ddot{u}(x)}{a\dot{u}^3(x)} & a\dot{u}^2(x) + \dfrac{1}{a}\left[\dot{v}(x) - \dfrac{y\ddot{u}(x)}{\dot{u}^2(x)}\right]^2 & 0 \\ 0 & 0 & a \end{bmatrix}, \quad (A1)$$

i.e., *real* and *symmetric* tensors, whose eigenvalues can be computed from the characteristic equation

$$(\Lambda - a)\left[\Lambda^2 + 1 - a\Omega(x,y)\right] = 0, \quad (A2)$$

where

$$\Omega(x,y) = \frac{1}{a^2\dot{u}^2(x)} + \dot{u}^2(x) + \frac{\left[-\dot{u}^2(x)\dot{v}(x) + y\ddot{u}(x)\right]^2}{a^2\dot{u}^4(x)}. \quad (A3)$$

Noting that $\Omega(x,y)$ in (A3) is a sum of squares, and hence *always positive*, Eq. (6) follows immediately from (A2) as a consequence of Descartes' rule of signs.

**Appendix B: Pertaining to (8)-(10)**

A general expression of the magnetic field in the presence of the transformation slab can be compactly written as a spectral integral

$$H_z(x,y) = \frac{1}{2\pi}\int_{-\infty}^{\infty} \hat{f}(k_y)\exp\left[i\varphi(k_x,k_y;x,y)\right]dk_y, \quad (B1)$$

where $k_x = \sqrt{\omega^2/c_0^2 - k_y^2}$, $\text{Im}(k_x) \geq 0$ is the *x*-domain wavenumber (with $c_0$ denoting the speed of light in vacuum),

$$\hat{f}(k_y) = \int_{-\infty}^{\infty} f(x)\exp(-ik_y y)dy \tag{B2}$$

is the plane-wave spectrum of the aperture field distribution in (7), and

$$\varphi(k_x,k_y;x,y) = \begin{cases} k_x(x-x_s)+k_y y, & x<-d, \\ k_x\left[au(x)-x_s\right]+k_y\left[v(x)+\dfrac{y}{\dot{u}(x)}\right]+\varphi_1(k_x,k_y), & |x|<d, \\ k_x x+k_y y+\varphi_2(k_x,k_y), & x>d. \end{cases} \tag{B3}$$

The phase function in (B3) is derived by solving the straightforward problem in the virtual space (plane-wave propagation in vacuum) and subsequently translating this solution in the actual physical space via the continuous coordinate transformation in (1), accounting for the possible phase displacements $\varphi_{1,2}(k_x,k_y)$ introduced by the reflectionless slab interfaces. Enforcing the field continuity conditions at these interfaces $(x=\mp d)$ yields, after straightforward algebra,

$$\varphi_1(k_x,k_y) = -k_x\left[d+au(-d)\right]-k_y v(-d), \tag{B4}$$

$$\varphi_2(k_x,k_y) = -k_x x_i - k_y y_0, \tag{B5}$$

with $x_i$ and $y_0$ defined in (9) and (10), respectively. Equation (8) then follows straightforwardly, by substituting (B4) [with (B5)] in (B1).

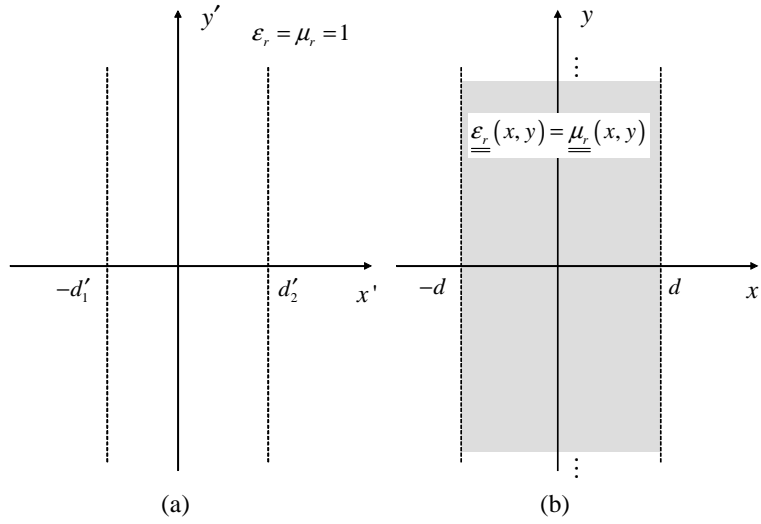

**Figure 1** – Problem geometry. The fictitious (vacuum) space $(x', y', z')$ is imaged via (1) [with (2)] into the slab region $|x| \le d$ (embedded in vacuum) in the actual physical space $(x, y, z)$, with the planar interfaces $x' = d'_{1,2} \equiv au(\mp d)$ imaged (apart from possible irrelevant rigid translations) as the planar interfaces $x = \mp d$. The curved-metrics-induced EM field effects are equivalently obtained in a flat, Cartesian space by filling-up the transformed slab region $|x| \le d$ with the transformation medium in (3).

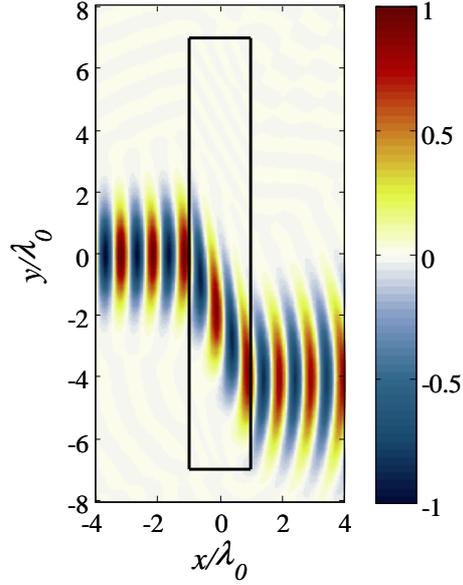

**Figure 2** – (Color online) An example known in the literature, but using the technique presented here: Geometry as in Fig. 1, with $d = \lambda_0$ (but truncated along the *y*-axis to an aperture of $14\lambda_0$), and transformation-slab in (13) with $\alpha = 2$ [i.e., $\theta_R = 60°$ in (15)], assuming a loss-tangent of $10^{-3}$. FEM-computed magnetic field map pertaining to a normally-incident collimated Gaussian beam (with waist of size $\sqrt{2}\lambda_0$, located at $x_s = -4\lambda_0$, i.e., $3\lambda_0$ away from the slab interface), displaying the lateral beam shift.

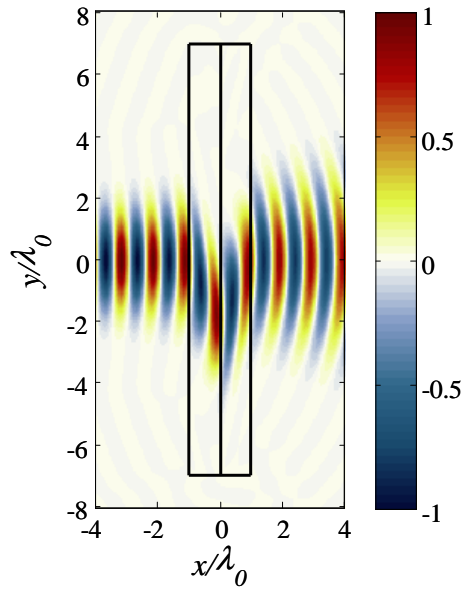

**Figure 3** – (Color online) As in Fig. 2, but for the perfect-radome (twin-crystal) configuration in (21), illustrating the beam-shift compensation.

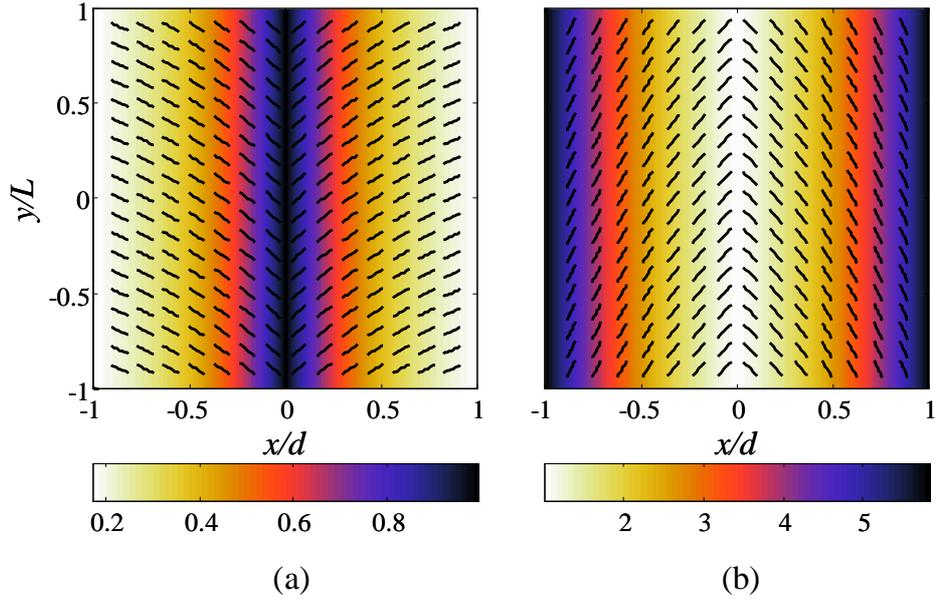

**Figure 4** – (Color online) Constitutive-parameter maps (transverse components) for the perfect-radome configuration in (22) (with $\beta=1$), shown in the principal reference system. As a reference, the principal axes directions $\xi$ and $\upsilon$ are shown as short segments, in (a) and (b), respectively.

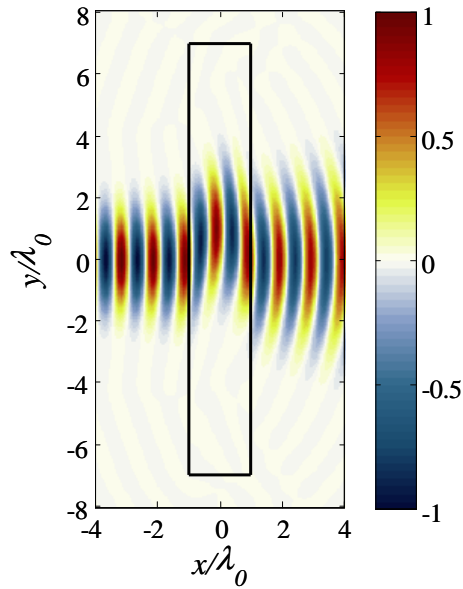

**Figure 5** – (Color online) As in Fig. 2, but for the perfect-radome configuration in (22) (cf. the constitutive parameters in Fig. 4).

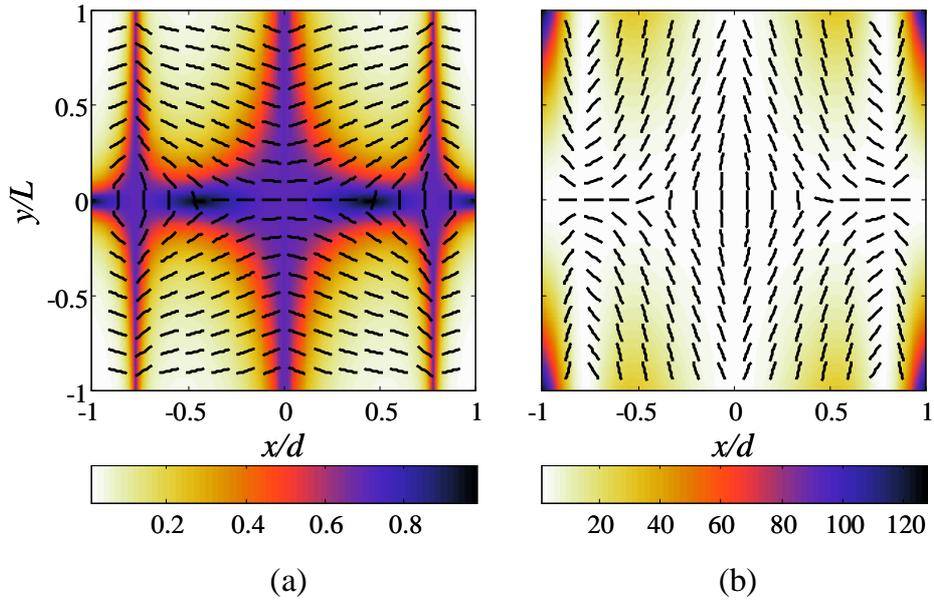

**Figure 6** – (Color online) As in Fig. 4, but for the configuration in (24) and (25) (with $n=2$ and $u_2 =1$).

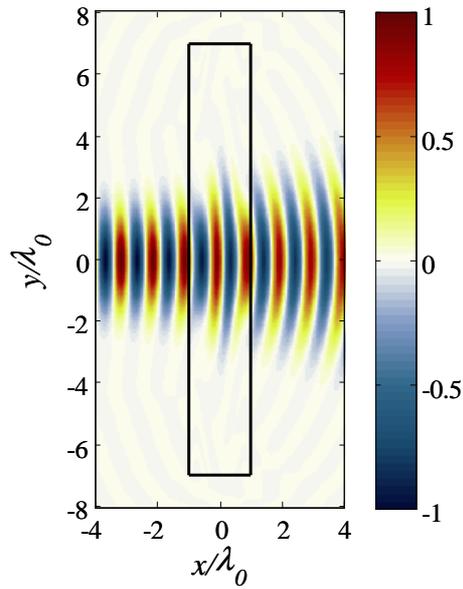

**Figure 7** – (Color online) As in Fig. 2, but for the perfect-radome configuration in (24) and (25) (cf. the constitutive parameters in Fig. 6).

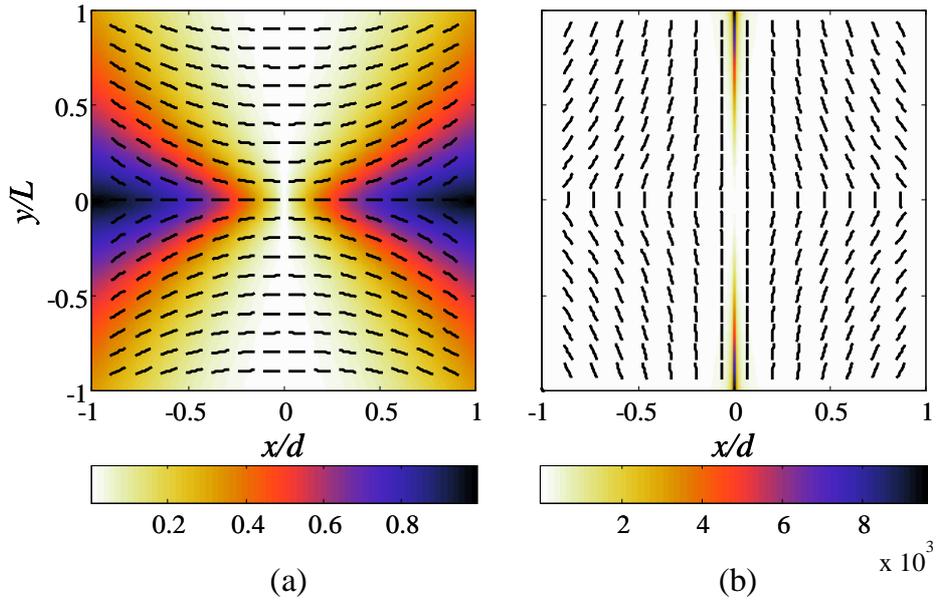

**Figure 8** – (Color online) As in Fig. 4, but for the configuration in (26) (with $\delta = d$ and $\Delta = d/100$).

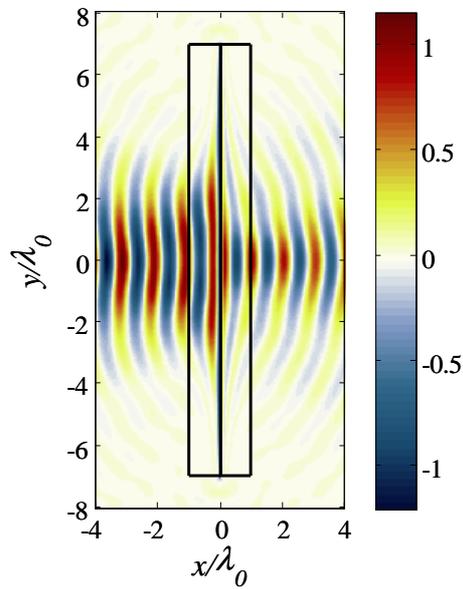

**Figure 9** – (Color online) As in Fig. 2, but for the configuration in (26) (cf. the constitutive parameters in Fig. 8).

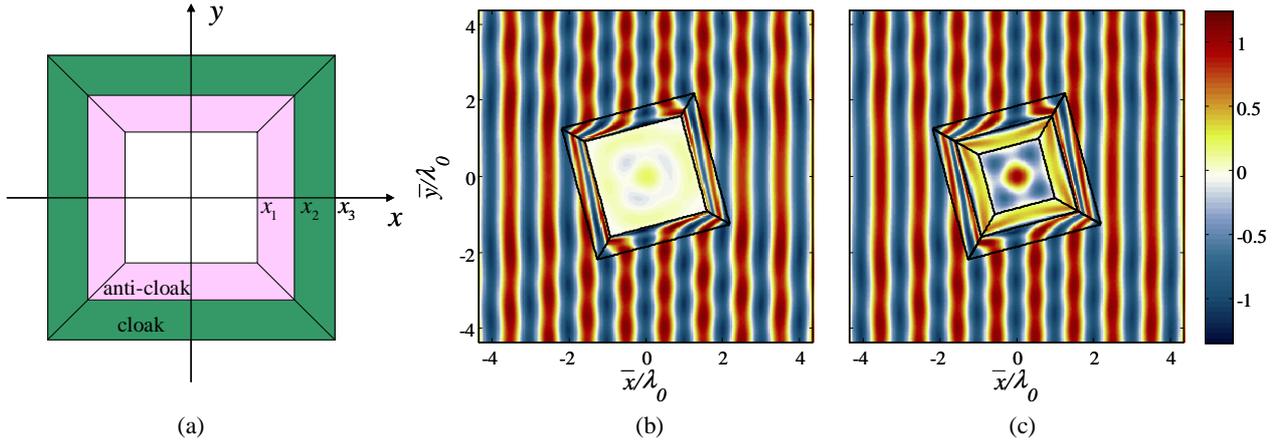

**Figure 10** – (Color online) (a) Square cloak/anti-cloak geometry [cf. (28) and (29)]. (b), (c) FEM-computed magnetic field maps pertaining to oblique (15°) plane-wave excitation in the presence and absence, respectively, of the anti-cloak shell (parameters: $x_1 = 0.81\lambda_0$, $x_2 = 1.30\lambda_0$, $x_3 = 1.79\lambda_0$, $\Delta_3 = x_1/100$, $\Delta_2 = 1.82\Delta_3$). Note that, for computational convenience, the oblique incidence is simulated using a 15°-rotated ($\bar{x}, \bar{y}$) coordinate system, where the illuminating wave impinges along the $\bar{x}$-axis.

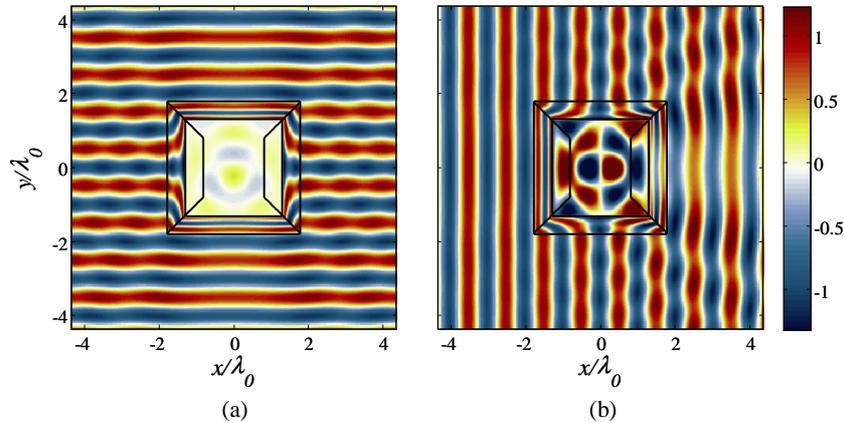

**Figure 11** – (Color online) As in Figs. 10(b) and 10(c), but with the two horizontal anti-cloak elements removed, and plane-wave incidence direction orthogonal (a) and parallel (b) to the corresponding sides.

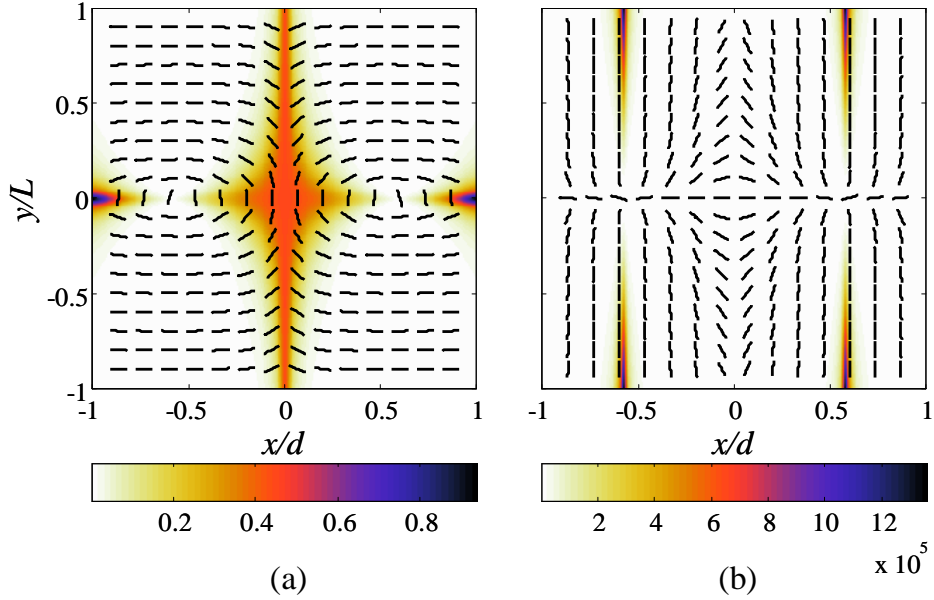

**Figure 12** – (Color online) As in Fig. 4, but for the DPS focusing configuration in (30) (with $x_i = 1.01d$ and $\Delta = d/10$).

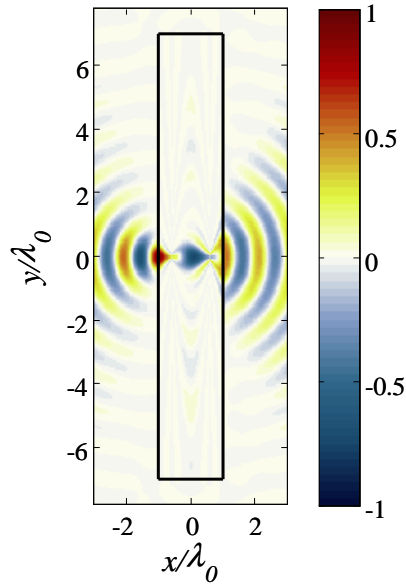

**Figure 13** – (Color online) As in Fig. 2, but for the DPS focusing configuration in (30) (cf. the constitutive parameters in Fig. 12), for a very weakly collimated Gaussian beam (with waist of size $\lambda_0/3$, located at $x_s = -1.01d$, i.e., $\lambda_0/100$ away from the slab interface), displaying the focusing effects at the image plane $x_i = -x_s$.